\documentstyle[12pt,aasms4]{article}
\def\ni{\noindent}
\def\spose#1{\hbox to 0pt{#1\hss}}
\def\lta{\mathrel{\spose{\lower 3pt\hbox{$\mathchar"218$}}
     \raise 2.0pt\hbox{$\mathchar"13C$}}}
\def\gta{\mathrel{\spose{\lower 3pt\hbox{$\mathchar"218$}}
     \raise 2.0pt\hbox{$\mathchar"13E$}}}

\begin{document}

\title{The Distribution of Burst Energy and Shock Parameters for Gamma-ray
Bursts}
\author{Pawan Kumar}
\affil{IAS, Princeton, NJ 08540}
\authoremail{pk@ias.edu}

\begin{abstract}
\baselineskip 15pt

We calculate the statistical distribution of observed afterglow flux, 
in some fixed observed frequency band, and at some fixed observer time 
after the explosion ($t_{obs}$) in two models -- one where the explosion 
takes place in a uniform density medium and the other where the 
surrounding medium has a power-law stratification such as is expected 
for a stellar wind. For photon energies greater than about 500 
electron-volt and $t_{obs}\gta 10^3$ sec the afterglow flux distribution 
functions for the uniform ISM and the wind models are nearly identical. 
We compare the width of the theoretical distribution with the observed 
x-ray afterglow flux and find that the FWHM of the distribution for 
energy in explosion and the fractional energy in electrons ($\epsilon_e$)
are each less than about one order of magnitude and the FWHM for the 
electron energy index is 0.6 or less.

\end{abstract}

\bigskip
\hskip0.3cm{\it Subject headings:\rm~ gamma rays: bursts -- gamma-rays: theory}

\vfill\eject
\baselineskip 15pt

\section{Introduction}

The improvement in the determination of the angular position 
of gamma-ray bursts (GRBs) by the Dutch-Italian satellite Beppo/SAX
has led to the discovery of extended emission in lower energy photons
lasting for days to months, which has revolutionized our understanding 
of GRBs (cf. Costa et al. 1997, van Paradijs et al. 1997, Bond 1997, 
Frail et al. 1997). The afterglow emission was predicted prior
to their actual discovery by a number of authors (Paczy\'nski \& Rhoads, 1993;
Meszaros \& Rees, 1993; Katz, 1994; Meszaros \& Rees 1997) based
on the calculation of synchrotron emission in a relativistic external 
shock. The afterglow observations have been found 
to be in good agreement with these theoretical predictions (cf. Sari, 1997;
Vietri, 1997a; Waxman, 1997; Wijers et al. 1997).

The medium surrounding the exploding object offers some clue as to
the nature of the explosion. Vietri (1997b), and Chevalier and Li (1999a-b)
in two very nice recent papers, have offered evidence that some
GRB afterglow light curves are best explained by a stratified 
circumstellar medium which suggests the death of a massive object 
as the underlying mechanism for gamma-ray burst explosions as was
suggested by Paczy\'nski (1998), and Woosley (1993). Possible 
further evidence in support of
such a model has come from the flattening and reddening of
afterglow emission, a few days after the burst, in optical
wavelength bands (eg. Bloom et al. 1999, Castro-Tirado \& Gorosabel 1999,
Reichart 1999, Galama et al. 1999).

The goal of this paper is to explore the afterglow flux in different
models, uniform ISM as well as stratified medium, and compare it with 
observations in a statistical sense, as opposed to comparison with
individual GRBs as carried out by Chevalier and Li (1999b). We will use 
this statistical comparison to constrain various physical parameters 
that determine the afterglow
luminosity such as the energy, $E$, the fractional energy in electrons
($\epsilon_e$) and magnetic field ($\epsilon_B$), the electron energy
index ($p$), and the circumstellar density $n$.

In the next section we discuss the afterglow flux and its distribution
and comparison with observations.

\section{Afterglow flux and its distribution}
\label{physical}

Consider an explosion which releases an equivalent of isotropic
energy $E$ in a medium where the density varies as $A r^{-s}$; $r$
is the distance from the center of the explosion, and $A$ is a
constant. The deacceleration radius, $r_d$, where the shell 
starts to slow down as a result of sweeping up the circumstellar 
material, and the deacceleration time, $T_{da}$, in the observer 
frame are given by

$$ R_{da} = \left[ {(17-4s) E\over 2\pi c^2 A \Gamma_0^2} \right]^{1/(3-s)},
   \quad\quad  T_{da} = {R_{da} \over 4\beta c (\Gamma_0/2)^2}, \eqno(1)$$
where $\Gamma_0$ is the initial Lorentz factor of the ejecta, and
$\beta\approx 1$ is a constant.

The time dependence of the shell radius and the Lorentz factor can
be obtained from the self-similar relativistic shock solution
given in Blandford \& McKee (1976)

$$ {R(t_{obs})\over R_{da} } \equiv X = t_1^{1/(4-s)}, \quad\quad
   \Gamma(t_{obs}) = {\Gamma_0\over 2} t_1^{-(3-s)/(8-2s)}, \quad\quad
   {\rm where} \quad t_1 \equiv {t_{obs}\over (1+z)T_{da}}, \eqno(2)$$

\ni and $t_{obs}$ is the time in observers' frame at redshift $z$. 

The magnetic field and the electron thermal Lorentz factor behind the
forward shock vary as,

$$ B = B_{da} \epsilon_B^{1/2} t_1^{-3/(8-2s)}, \quad\quad
   \gamma_e = \epsilon_e \left({m_p\over m_e}\right) {\Gamma\over 2^{1/2}},
\quad\quad {\rm where} \quad\quad
B_{da} = \left[ {2(17-4s) E\over \Gamma_0^2 R_{da}^3} \right]^{1/2}, 
\eqno(3)$$
and $\epsilon_B$ and $\epsilon_e$ are the fractional energies in the
magnetic field and the electrons, respectively.

Using these results we find that the peak of the synchrotron frequency
($\nu_m$) and the cooling frequency ($\nu_c$), in the observer frame, are

$$ \nu_m = \nu_{m,da} \epsilon_e^2 \epsilon_B^{1/2} t_1^{-3/2}, \quad\quad 
   \nu_c = \nu_{c,da} \epsilon_B^{-3/2} t_1^{(3s-4)/(8-2s)},  \eqno(4)$$
where
$$ \nu_{m,da} = {1\over 32\pi 2^{1/2}} {q B_{da} m_p^2\over c m_e^3},
    \quad\quad \nu_{c,da} = {9\pi\over 4 2^{1/2}} {m_e q c^3 \Gamma_0^3\over
                   \sigma_T^2 B_{da}^3 R_{da}^2}. \eqno(5)$$

The synchrotron self-absorption frequency (in the observer frame) is given by

$$ \nu_A = \left[ {27^{1/2} m_e c^2 \sigma_T A R_{da}^{1-s} B_{da}
                    \Gamma_0\over 64\pi q m_p^2} \right]^{3/5}
            \epsilon_e^{-3/5} \epsilon_B^{3/10} t_1^{-0.3(4+s)/(4-s)}
            \left[ \min(\nu_m, \nu_c)\right]^{-1/5}. \eqno(6)$$

The energy flux at the peak of the spectrum is given by

$$ f_{\nu_p} = {27^{1/2}\over 32\pi} \left[{m_e\sigma_T c^2\over 
   q m_p d_L^2} \right] A\Gamma_0 B_{da} R_{da}^{(3-s)} \epsilon_B^{1/2}
    (1+z) t_1^{-s/(8-2s)}, \eqno(7)$$
\ni where $\nu_p=\min\{\nu_m,\nu_c\}$ i.e. for $\nu_m>\nu_c$ the peak occurs 
at $\nu_c$ instead of at $\nu_m$.

The equations for $s=2$ are as in Chevalier and Li (1999a-b)
and are given here for easy reference. The flux at an arbitrary 
observed frequency $\nu$ can be calculated following
Sari, Piran and Narayan (1998) in terms of $f_{\nu_p}$, $\nu_m$,
$\nu_c$ and $\nu_A$. For the particularly important case of 
$\nu$ greater than $\nu_m$ and $\nu_c$ the observed flux is:

$$ f_\nu = f_{\nu_p} \nu_c^{1/2} \nu_m^{(p-1)/2} \nu^{-p/2} =
   {3^{2.5} c^5\over \nu^{p/2} d_L^2} \left({q m_p^2\over m_e^3} 
   \right)^{(p-2)/2} {\epsilon_e^{p-1} \over \epsilon_B^{(2-p)/4} }
   {(1+z)^{(3p+2)/4}\over t_{obs}^{(3p-2)/4} }
   \left[ { (17-4s) E\over 2^{10} \pi^2 c^5}\right]^{(p+2)/4}, \eqno(8)$$

\ni Note that the flux does not depend on the circumstellar density 
parameters $A$ and $s$ when $\nu>\nu_c$, except through an unimportant 
multiplicative factor $(17-4s)^{(p+2)/4}$.

The frequencies $\nu_m/(\epsilon_e^2\epsilon_B^{1/2})$, $\nu_c\epsilon_B^{3/2}$
and $\nu_A$ are shown in figure 1. The afterglow flux for the uniform
ISM and the wind models only differ when $\nu<\nu_c$. Since
the cooling frequency decreases with time for the uniform ISM model
and increases with time for the wind model, one of the
best ways to distinguish between these models is by observing the
behavior of the light-curve when $\nu_c$ crosses the observed
frequency band as has been pointed out by Chevalier \& Li (1999).
The predictions and comparison of individual GRB
lightcurves for the two models will be discussed in some detail
in a separate paper. Here we turn our attention to the statistical 
property of the afterglow light curve in the two models.

\bigskip
\subsection{Afterglow flux distribution function}
\medskip

The distribution function for GRB afterglow flux, $P_t(L_\nu)$, at a 
frequency $\nu$, and time $t_{obs,g}$ is the probability that the afterglow
luminosity (isotropic) is $L_\nu$ at time $t_{obs,g}$ after the explosion;
$\nu$, $t_{obs,g}$ and $L_\nu$ are measured in the rest frame of the host 
galaxy.

The width of $P_t(L_\nu)$ is a function of the width of the distribution
function for $E$, $\epsilon_e$, $\epsilon_B$, $A$ and $p$. Assuming that
all these variables are independent Gaussian random variables the
standard deviation (SD) for $\log(L_\nu)$, $\sigma_{L_\nu}$, can be obtained 
from equation (8), when $\nu>\nu_c$ \& $\nu_m$, and is given by

$$\sigma^2_{L_\nu} = \left({p+2\over 4}\right)^2\sigma^2_E
   + (p-1)^2\sigma^2_{\epsilon_e} + \eta\sigma^2_p + \left({p-2\over 4}\right)^2
     \sigma^2_{\epsilon_B}, \eqno(9)$$
where 
$$ \eta = {1\over 16} \left[ 2\log{q m_p^2\over m_e^3} + \log\left(
    { 17\bar\epsilon_B\bar\epsilon_e^4\bar E\over 2^{10}\pi^2
    c^5 \nu^2 t_{obs,g}^3} \right) \right]^2, \eqno(10)$$
$\bar\epsilon$ and $\bar E$ are the mean values of $\epsilon$ and $E$,
and $\sigma_E$, $\sigma_{\epsilon_e}$, $\sigma_{\epsilon_B}$ and $\sigma_p$
are the standard deviation for $\log E$, $\log\epsilon_e$, $\log\epsilon_B$
and $p$ respectively; $\eta$ for x-ray (10 kev) and optical (2.5 ev) photons
are shown in figure 2.

The standard deviation of the flux in the 2-10 kev band at 5 hours after
the burst ($\sigma_{L_\nu}$) is approximately 0.58 (Kumar \& Piran, 1999). 
Fig 1 shows that this energy band is above $\nu_c$ \& $\nu_m$ so long
as $\epsilon_B>10^{-4}$ and the density of the surrounding medium is not
too small. Moreover, $\eta=5$ for this energy band (see fig. 2), from 
which we obtain an upper limit on $\sigma_p$ of 0.26 and the full-with 
at half-maximum of the distribution for $p$ to be less than about 0.6.
We note that the electron energy index $p$ lies between 2 and 3 for 
supernovae remanents (cf. Chevalier 1990, Weiler et al. 1986), and
Chevalier and Li (1999b) point out that the range in $p$ for GRB 
afterglows is at least $\sim$ 2.1--2.5.

We can use the variation of $\eta$ with time or $\nu$ (see figs. 2 \& 3) 
to obtain $\sigma_p$ from the observed variation to the width of 
$\sigma_{L_\nu}$. This is of course equivalent to the determination 
of $p$ from the slope of the light curve or the spectral slope.
The currently available data, consisting of 7 bursts with known redshifts,
does not provide a good constraint on the temporal variation of 
$\sigma_{L_\nu}$, however, HETE II and Swift would increase the GRB 
afterglow data base by more than an order of magnitude and provide 
much better constraint on $\sigma_p$. 

Unless $\epsilon_B$ varies by many orders of magnitude from one burst
to another the last term in equation (9) is very small and can be neglected.
Equating the first two terms individually to $\sigma_{L_\nu}=0.58$ we 
obtain $\sigma_E< 0.51$, and $\sigma_{\epsilon_e}<0.39$ (for $p=2.5$). 
For comparison, if the first three terms in equation (9) were to contribute 
equally to $\sigma_{L_\nu}$ then we obtain $\sigma_E=0.29$, 
$\sigma_{\epsilon_e} = 0.22$, and $\sigma_p=0.15$. The mean values for $E$ 
and $\epsilon_e$ are not well determined by this procedure (but see the 
discussion below), however $\bar E \bar\epsilon_e^{^{4(p-1)/(p+2)}}$ can be 
accurately obtained from the observed distribution and is $\approx 10^{52}$ erg.

So far we have discussed a nearly model independent procedure for 
determining $\sigma_p$ and a linear combination of $\sigma^2_E$ and 
$\sigma^2_{\epsilon_e}$ that relies on making observations in a frequency
band that lies above $\nu_m$ \& $\nu_c$. In order to determine $\sigma_E$ and
$\sigma_{\epsilon_e}$ separately we need to have some knowledge of $\nu_c$
and $\nu_m$, and therefore the result is model dependent and less certain.
For instance, if the cooling and the peak synchrotron frequencies are
known at some time, even if only approximately, then $\sigma_E$ can be 
determined from the distribution of the observed flux at a frequency $\nu$ 
such that $\nu_c<\nu<\nu_m$. The SD for $f_1\equiv L_\nu t_{obs,g}^{1/4}
\nu^{1/2}$ at such an intermediate frequency is independent of $s$ and
is given by:

$$ \sigma^2_{f_1} = {9\over 16}\sigma^2_E + {1\over 16}\sigma^2_{\epsilon_B}
     \approx {9\over 16}\sigma^2_E. \eqno(11)$$
Once $\sigma_E$ is known, equation (9) can be used to determine 
$\sigma_{\epsilon_e}$. Observations at low frequencies i.e. $\nu<\nu_m, \nu_c$,
can be used to constrain $\sigma_{\epsilon_B}$ and $\sigma_A$ which when
combined with the flux at the peak of the spectrum could be used to
determine $\sigma_{\epsilon_B}$ \& $\sigma_A$ separately with the use
of the following equations,

$$ \sigma^2_{f_2} = {4\over (4-s)^2}\sigma_A^2 + {4\over 9}\left( 
    \sigma^2_{\epsilon_e} + \sigma^2_{\epsilon_B} \right) + \left(
   {14-5s\over 12-3s}\right)^2 \sigma^2_E,  \quad\quad {\rm for}\quad
    \nu<\nu_m<\nu_c, \eqno(12)$$

$$ \sigma^2_{f_3} = {4\over 9(4-s)^2}\sigma_A^2 + \sigma^2_{\epsilon_B} + 
    \left( {14-6s\over 12-3s}\right)^2 \sigma^2_E, \quad\quad\quad\quad\quad 
   \quad{\rm for}\quad \nu<\nu_c<\nu_m, \eqno(13)$$
and
$$ \sigma^2_{f_4} = {1\over 4}\sigma^2_{\epsilon_B} + {4\over (4-s)^2}\sigma^2_A
    + \left( {8-3s\over 8-2s}\right)^2 \sigma^2_E,  
{\hskip 5.9cm} \eqno(14)$$
where $f_1\equiv \nu^{-1/3} L_\nu t_{obs,g}^{(s-2)/(4-s)}$, $f_2\equiv 
\nu^{-1/3} L_\nu t_{obs,g}^{(3s-2)/(12-3s)}$, $f_3\equiv L_{\nu_p} 
t_{obs,g}^{s/(8-2s)}$, and $L_{\nu_p}$ is the isotropic luminosity at 
the peak of the spectrum.

Another approach to determining the burst and shock parameters is to
compare the observed flux distributions at several different frequencies
and time with the theoretically calculated distributions. The latter can
be easily calculated by varying $E$, $\epsilon_e$, $\epsilon_B$, $A$ and $p$ 
randomly and solving for the flux using the equations given in the last
section. Fig. 3 shows a few cases of flux distribution functions for 
several different $\nu$ and $t_{obs}$. The advantage of this procedure is
that it does not require observational determination of various characteristic
frequencies i.e. $\nu_m$, $\nu_c$, and $\nu_A$, which is difficult to do unless
we have good spectral and temporal coverage over many orders of magnitude.
Moreover, this procedure can be used even without the knowledge of burst
redshifts; the redshift distribution of a representative sub-sample of 
bursts can be used to calculate the expected theoretical distribution of 
observed afterglow flux which can be directly compared with the observed
flux distribution to yield the burst and shock parameters.

It should be noted that if $\nu_m$, $\nu_c$, $\nu_A$ and $L_{\nu_p}$
can be determined accurately then $E$, $\epsilon_e$, $\epsilon_B$, and
$A$ can be obtained for individual bursts as described in eg. Chevalier \& Li 
(1999b), and Wijers \& Galama (1999), and there is no need to resort to 
the statistical treatment discussed above.

\bigskip
\section{Conclusion}
\medskip

We have described how the distribution function for GRB afterglow
flux can be used to determine the width of the distribution
function for the energy in the explosion and the shock parameters
such as $\epsilon_e$, $\epsilon_B$ (the fractional energy in
electrons and the magnetic field), $p$ (the power law index for
electron energy), and $A$ (the interstellar density parameter).

The afterglow flux at a frequency above the cooling and the synchrotron
peak frequencies is independent of interstellar density and scales as 
$E^{(p+2)/4}\epsilon_e^{(p-1)}\epsilon_B^{(p-2)/4}$ for uniform ISM
as well as for energy deposited in a stellar wind with power-law
density stratification. Using the flux in 2-10 kev band for 7 bursts with
known redshifts, 5 hours after the burst, which meet the above criteria,
we find that the full width at half maximum of the distribution for 
Log$E$ is less than 1.2, for Log$\epsilon_e$ is less than 0.9 and 
$p$ is less than 0.6. The width for the distribution of $p$ is 
consistent with the range in $p$ deduced from afterglow emissions 
(cf. Chevalier \& Li, 1999b). The width of the distribution for $p$ can 
be more accurately determined from the time variation, or frequency 
dependence, of the width of the afterglow flux distribution. 

For a more accurate determination of the distribution of other 
parameters we need to determine the cooling and the synchrotron 
peak frequencies (at least approximately), or otherwise compare 
the theoretical and the observed distributions 
for flux at several different frequencies covering the range above 
and below the cooling and the synchrotron frequencies. 

HETE II and Swift missions are expected to significantly increase 
the number of GRBs with observed afterglow emission and should therefore 
provide a more accurate determination of afterglow luminosity function 
and the distribution for burst energy and shock parameters.

\medskip
\ni {\bf Acknowledgment:} I am indebted to Roger Chevalier for 
many useful discussions and for clarifying several points. I thank Alin 
Panaitescu, Tsvi Piran and Bohdan Paczy\'nski for numerous exciting
discussions about gamma-ray bursts, and Eliot Quataert for comments
on the paper. I thank E. Waxman for sending me his recent preprint 
which bears some technical similarity with this work, although the 
results and conclusions are different.

\vfill\eject

\begin{figure}
\plotone{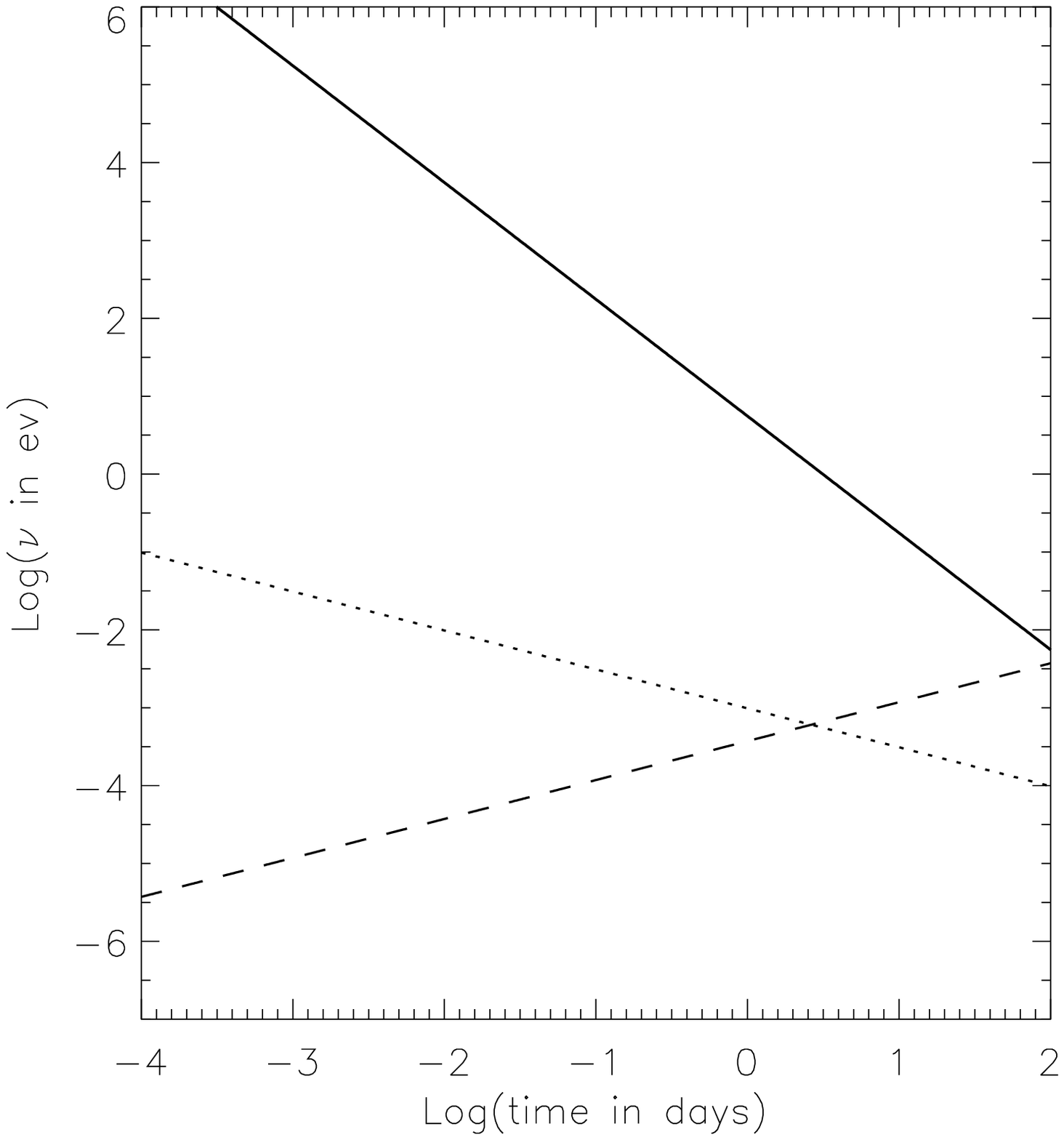}
\figcaption{\baselineskip=14pt 
The figure shows several different frequencies, in electron
volts, as a function time in observer frame. The solid line is the
peak synchrotron frequency for $\epsilon_B=1$ \& $\epsilon_e=1$ for the
wind model i.e.  $\nu_m/(\epsilon_B^{1/2}\epsilon_e^2)$ for $s=0$; $\nu_m$ 
for uniform interstellar medium is larger by a constant factor of 1.374.
The dotted curve is the cooling frequency, $\nu_c\epsilon_B^{3/2} n_0$,
for uniform ISM ($n_0$ is particle number density in the ISM). The
dashed curve is $\nu_c\epsilon_B^{3/2} A_*^2$ for the wind model ($A_*$
is the baryon density in the wind in the units of 5x10$^{11}$ g cm$^{-1}$).
The energy in the explosion $E$ has been taken to be 10$^{52}$ erg
and $p=2.5$.
\label{fig:fig1}
}\end{figure}

\begin{figure}
\plotone{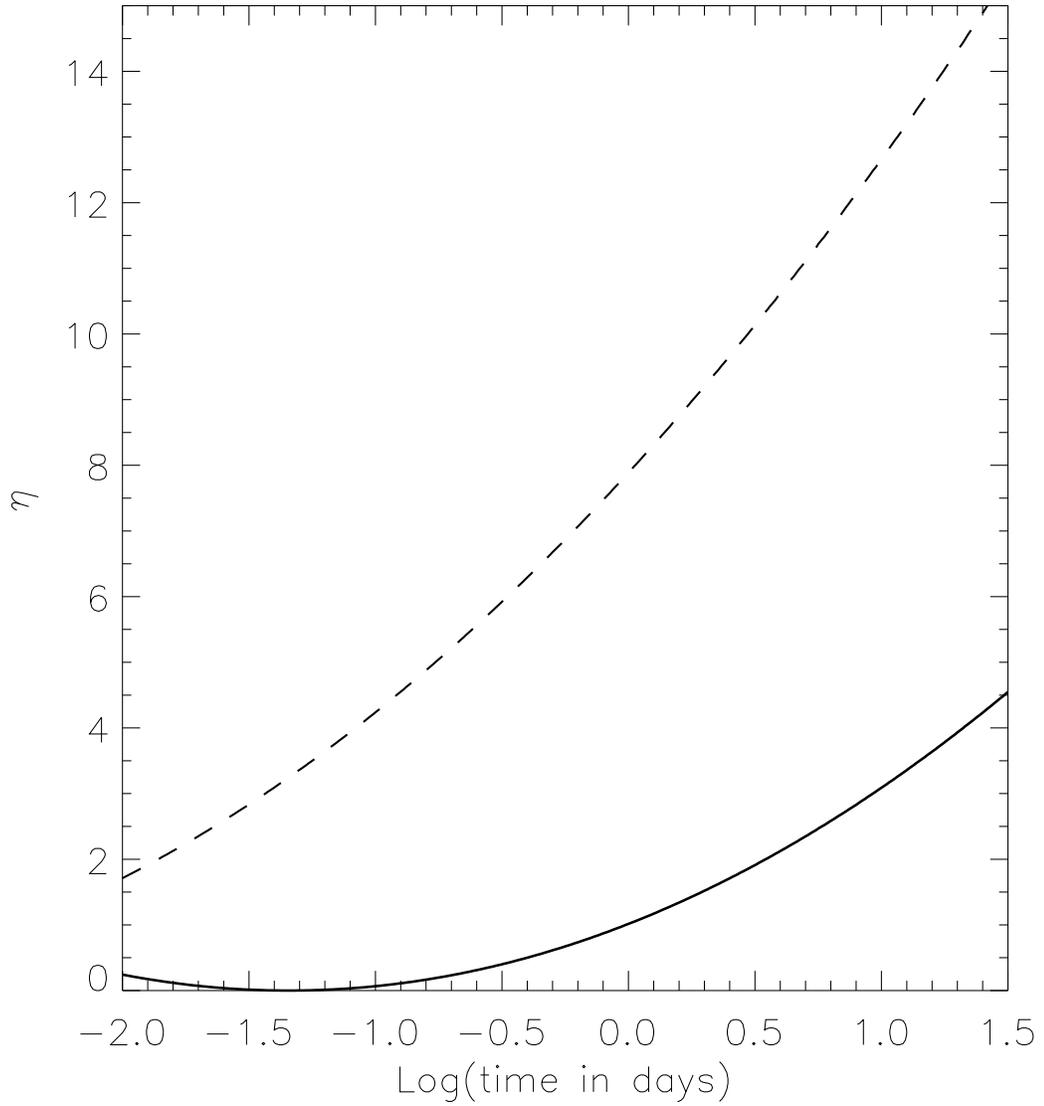}
\figcaption{\baselineskip=13pt 
$\eta$ as defined in equation (10) is shown as a function
of time for two different frequencies --- 2.5 ev (solid curve)
and 10 kev (dashed curve). We took $E=10^{52}$ erg, $\bar\epsilon_e=0.1$,
and $\bar\epsilon_B=0.03$ for these calculations.
\label{fig:fig2}
}\end{figure}

\begin{figure}
\plotone{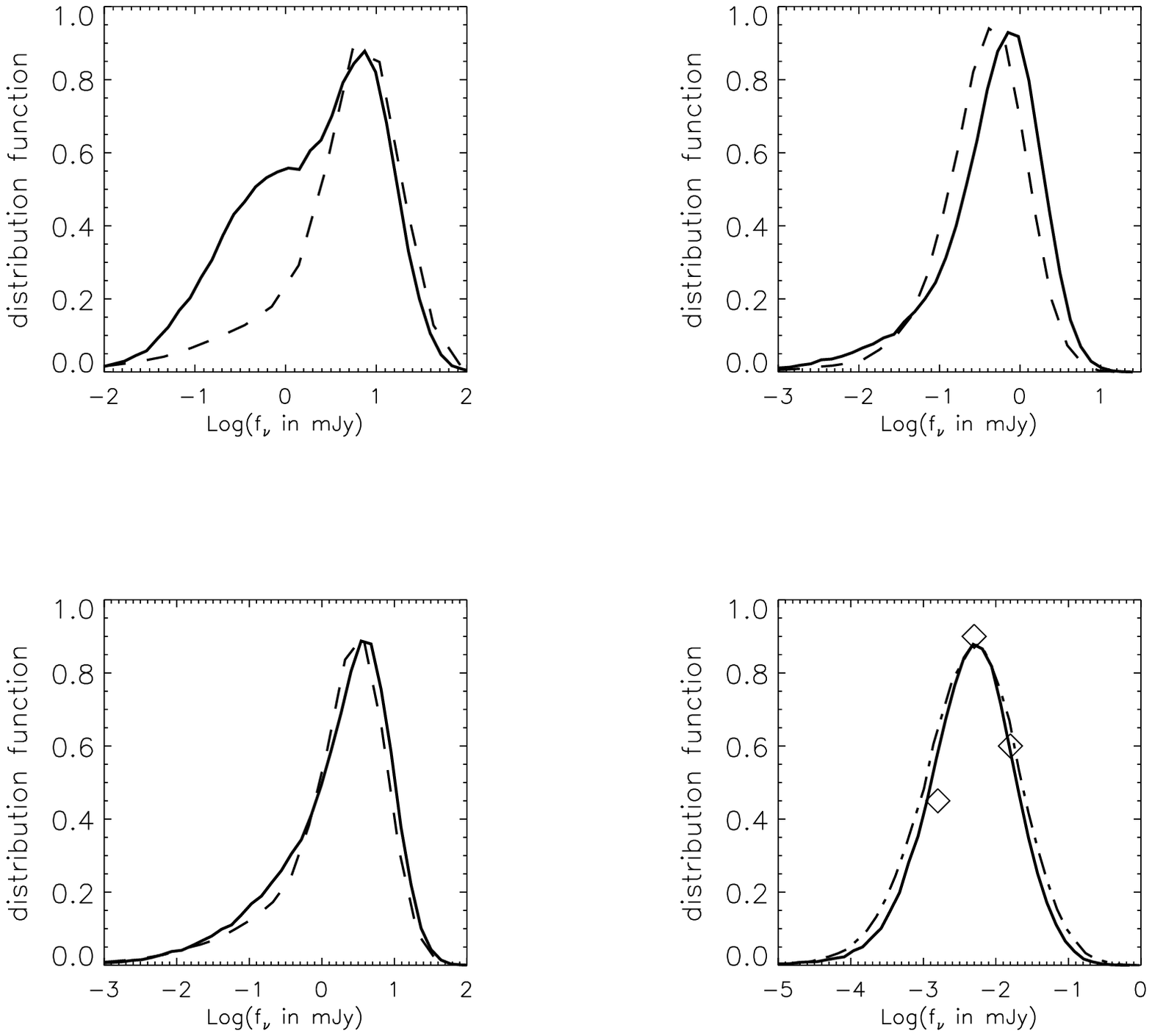}
\figcaption{\baselineskip=13pt
Distribution function for afterglow flux per unit frequency at photon 
energy of 1.0 ev, 0.1 hr after the burst (top left panel), for uniform ISM
(solid line) and stratified medium with $s=2$ (dashed line).
The top right panel shows the DFs at photon energy of 100 ev and at 0.1 hr
after the burst; solid line is for ISM and dashed line is for $s=2$. 
The lower left panel shows DFs for photon energy of 1 ev at 1 hr
after the burst; solid line is for ISM and dashed line is for $s=2$.
The lower right panel shows the DFs for 10 kev energy photons at 0.1 hr 
after the burst (solid line) and at 10 hr (dash-dot curve) for the uniform 
ISM model; the dash-dot curve has been shifted to the right by 2.4.
The DFs for the wind model ($s=2$) for 10 kev energy band are identical
to the uniform ISM DFs at times greater than 0.5 hr except for an 
overall shift of the curve to lower energy by a factor of $\sim 2$. 
For all of these calculations the FWHM of Log$E$, Log$\epsilon_e$, 
Log$\epsilon_B$, $A$ and $p$ was taken to be 0.8, 0.7, 3.0, 2.4, and 0.25
respectively, and the burst redshift was taken to be 1.5.
Moreover, $\bar E=10^{52}$erg, $\bar\epsilon_e=0.1$, $\bar\epsilon_B=0.01$, 
and $\bar p= 2.3$. The diamonds in the lower panel are for bursts
with known redshifts; the error bar associated with each point is
much larger than the size of the symbol.
\label{fig:fig3}
}\end{figure}


\begin{references}
\bigskip

\ni Blandford, R.D., and McKee, C.F., 1976, Phys. Fluids, 19, 1130

\ni Bloom, J.S., et al., 1999, Nature 401, 453

\ni Bond, H.E. 1997, IAU Circ. 6654

\ni Catro-Tirado, A., and Gorosabel, J., 1999, A\&AS 138, 449

\ni Chevalier, R.A., 1990, in Supernovae, A.G. Petschek (ed.), Springer-Verlag

\ni Chevalier, R.A., and Li, Z-Y, 1999a, 520, L29

\ni Chevalier, R.A., and Li, Z-Y, 1999b, Astro-ph/9908272

\ni Costa, E., et al. 1997, IAU Circ. 6572

\ni Frail, D.A. et al. 1997, Nature 389, 261

\ni Galama, T.J., et al., 1999, astro-ph/9907264

\ni Katz, J.I., 1994, ApJ 432, L107

\ni M\'esz\'aros, P., and Rees, M.J., 1993, ApJ 405, 278

\ni M\'esz\'aros, P., and Rees, M.J., 1997, ApJ 476, 232

\ni Paczy\'nski, B., 1998, ApJ 494, L45

\ni Paczy\'nski, B., and Rhoads, J., 1993, ApJ 418, L5

\ni Reichart, D.E., 1999, ApJ 521, L111

\ni Sari, R., 1997, ApJ 489, L37

\ni Sari, R., \& Piran, T. \& Narayan, R. 1998, ApJ 497, L17

\ni van Paradijs, J., et. al. 1997, Nature 386, 686

\ni Vietri, M., 1997a, ApJ 488, L33

\ni Vietri, M., 1997b, ApJ 488, L105

\ni Waxman, E., 1997, ApJ 489, L33

\ni Weiler, K.W., Sramek, R.A., Panagia, N. van der Hulst, J.M., and
    Salvati, M., 1986, ApJ 301, 790

\ni Wijers, R.A.M.J., and Galama, T.J., 1999, ApJ 523, 177

\ni Wijers, R.A.M.J., Rees, M.J., \& M\'esz\'aros, P., 1997, MNRAS 288, 51

\ni Woosley, S.E., 1993, ApJ 405, 273

\end{references}
\end{document}